\documentclass[sigconf]{acmart}

\usepackage{graphicx,subfigure}
\usepackage{caption}
\usepackage{import}
\usepackage{comment}
\usepackage{tikz}
\usepackage{amsmath}
\usepackage{MnSymbol}
\usepackage{booktabs}
\usepackage[linesnumbered,ruled,vlined]{algorithm2e}
\SetKwInput{KwInput}{Input}               
\SetKwInput{KwOutput}{Output}
\usepackage{xurl}
\usepackage{multirow}
\usepackage{tablefootnote}

\begin{document}


\title{DF-SCA: Dynamic Frequency Side Channel Attacks are Practical}

\author{Debopriya Roy Dipta}
\affiliation{%
  \institution{Iowa State University}
  \streetaddress{}
  \city{Ames}
  \state{Iowa}
  \country{USA}
  \postcode{50010}
}
\email{roydipta@iastate.edu}

\author{Berk Gulmezoglu}
\affiliation{%
  \institution{Iowa State University}
  \streetaddress{}
  \city{Ames}
  \state{Iowa}
  \country{USA}
  \postcode{50010}
}
\email{bgulmez@iastate.edu}

\begin{abstract}

The arm race between hardware security engineers and side-channel researchers has become more competitive with more sophisticated attacks and defenses in the last decade. While modern hardware features improve the system performance significantly, they may create new attack surfaces for malicious people to extract sensitive information about users without physical access to the victim device. Although many previously exploited hardware and OS features were patched by OS developers and chip vendors, any feature that is accessible from userspace applications can be exploited to perform software-based side-channel attacks.

In this paper, we present DF-SCA, which is a software-based dynamic frequency side-channel attack on Linux and Android OS devices. We exploit unprivileged access to \textit{cpufreq} interface that exposes real-time CPU core frequency values directly correlated with the system utilization, creating a reliable side-channel for attackers. We show that Dynamic Voltage and Frequency Scaling (DVFS) feature in modern systems can be utilized to perform website fingerprinting attacks for Google Chrome and Tor browsers on modern Intel, AMD, and ARM architectures. We further extend our analysis to a wide selection of scaling governors on Intel and AMD CPUs, verifying that all scaling governors provide enough information on the visited web page. Moreover, we extract properties of keystroke patterns on frequency readings, that leads to 95\% accuracy to distinguish the keystrokes from other activities on Android phones. We leverage inter-keystroke timings of a user by training a k-th nearest neighbor model, which achieves 88\% password recovery rate in the first guess on Bank of America application. Finally, we propose several countermeasures to mask the user activity to mitigate DF-SCA on Linux-based systems.

\end{abstract}

\settopmatter{printfolios=true}

\maketitle

\section{Introduction}

The computation power of modern chips is one of the primary competitions among chip vendors, which constantly evolves in parallel with new chip generations. In general, computation capabilities of chips are improved with several techniques such as higher pipeline bandwidth, increasing number of computational units, improving clock frequency, etc. Due to the increasing popularity of mobile devices such as laptops and mobile phones in early 2000s, it has become more important for Intel~\cite{intel_cpu_efficiency}, ARM~\cite{arm_cpu_efficiency}, and AMD~\cite{amd_cpu_efficiency} to improve the battery life by developing more energy efficient CPUs. The same effort is also required to lower the energy consumption of CPUs in cloud computing and large-scale data centers that are responsible for an estimated 1.8\% of US electricity consumption~\cite{cloud_power}. 

The common mechanism to decrease the power consumption of a CPU is to adjust the energy consumption by integrating a sophisticated dynamic voltage supply into the system. This technology is known as Dynamic Voltage and Frequency Scaling (DVFS)~\cite{semeraro2002energy,le2010dynamic} that is responsible of adjusting the current voltage dynamically as well as the CPU frequency based on the system workload. This approach significantly reduces the energy consumption of CPUs as most of the time CPUs do not need to operate at the peak performance. DVFS technology is introduced in several CPU families under different names such as Enhanced Intel SpeedStep~\cite{intel2004speedstep}, AMD Powernow!~\cite{amd_powernow}, ARM Intelligent Energy Management~\cite{arm_iem}. Since the amount of energy consumption is decreased significantly with DVFS, they are also adapted to network devices~\cite{soteriou2007software}, hard drives~\cite{zhu2005hibernator}, and memory modules~\cite{fan2003synergy}.

Even though several hardware and OS-based features improve the computation and energy efficiency of the system, these features may be exploited to violate the privacy and security of the users. Several OS-based features such as performance counters~\cite{gulmezoglu2017perfweb}, memory API~\cite{naghibijouybari2018rendered}, Intel RAPL~\cite{lipp2021platypus}, data-usage statistics~\cite{spreitzer2016exploiting}, and system interrupts~\cite{zhang2009peeping} have been leveraged to collect user-specific data such as passwords, cryptographic keys, visited websites, and so on. All these attacks show that OS-based sensors that are accessible from user-space applications could be a source for software-based side-channel attacks, that could be of utmost importance to mitigate for OS and hardware developers. Although read access to these files is mostly restricted from userspace applications, new attack resources can still be discovered.

In this paper, we consider a threat scenario in which a malicious userspace application aims to detect the visited websites in browsers as well as user entered passwords in sensitive applications. For this purpose, DF-SCA utilizes \textit{cpufreq} directory in Linux and Android systems to monitor real-time dynamic CPU frequency, which is available to all userspace applications. Even though this feature has been exploited in the context of covert channels~\cite{alagappan2017dfs,miedl2018frequency,kalmbach2020turbocc} and cryptographic attacks~\cite{wan2022hertzbleed} in several platforms, it has not been investigated whether dynamic frequency readings can be leveraged to infer the user activity in a system by a malicious application. We observe that there two main challenges that prevent the usage of frequency measurements as an efficient side-channel attack: i) System noise introduced by the background processes affects the CPU frequency, leading to noisy measurements, ii) the sampling rate of frequency monitoring in \textit{cpufreq} directory is lower than other exploited OS and hardware features~\cite{gulmezoglu2017perfweb,naghibijouybari2018rendered,zhang2021red}. Despite these challenges we show that coarse-grain and noisy measurements still provide sufficient information to perform website fingerprinting, keystroke detection, and password recovery attacks with the utilization of Machine Learning and Deep Learning algorithms. 

\noindent\textbf{Our Contribution:}
In this paper, we show that
\begin{itemize}
    \item Low resolution dynamic frequency readings through Linux \textit{cpufreq} interface provide sufficiently-detailed information on the user activity on Intel, AMD, and ARM architectures.
    \item The collected frequency values can be used to train multi-class ML models to distinguish visited websites on Google Chrome and Tor browsers with 97\% and 73\% success rates, respectively. The experiments are carried out on Intel, AMD, and ARM devices to prove its viability over different micro-architectures.
    \item Website fingerprinting attacks are applicable for all available scaling governors on Intel and AMD architectures.
    \item User keystrokes on Android mobile phones leave a fingerprint on CPU frequency readings that leads to 95\% accuracy to capture the correct number of keystrokes.
    \item Inter-keystroke timings can be analyzed with ML algorithms to detect the entered passwords on the Bank of America application with 88\% accuracy.
    \item A universal ML model can be trained with collected website fingerprints from different architectures that can reach up to 92.3\% classification rate.
\end{itemize}


\noindent\textbf{Outline:} The rest of the paper is organized as follows: Section \ref{sec:background} provides background on DVFS. Section \ref{sec:threat_model} explains the threat model for DF-SCA. Section \ref{sec:exp_setup} gives information about the experiment setup. Section \ref{sec:case1} demonstrates the website fingerprinting attacks on Intel, AMD, and ARM architectures. Section \ref{sec:case2_password_detection} shows the applicability of DF-SCA on Android devices for keystroke detection. Section \ref{sec:related_work} provides an overview on previous studies related to website fingerprinting and keystroke detection. Section \ref{sec:countermeasures} proposes several countermeasures to prevent DF-SCA on modern devices. Section~\ref{sec:discussion} explores the scope of website fingerprinting  with different scaling governors and a universal ML-based model. Finally, Section \ref{sec:conclusion} concludes the study.
\section{Background}\label{sec:background}

In this section, we give an insight into dynamic voltage and frequency scaling (DVFS) that is utilized to perform the website detection and password exploitation attacks on Intel, ARM, and AMD architectures.

\subsection{Dynamic Voltage and Frequency Scaling (DVFS)}\label{subsec:dfs}

Modern processors are capable of operating with different voltage configurations and clock frequency, referred to as P-states or Operating Performance Points. The number of executed instructions in a given time can be increased with higher clock frequency and voltage while more energy is consumed over time by the CPU in the current P-state. Thereby, there is a natural trade-off between the energy consumption and the current CPU usage. As the CPU utilization is not at 100\% all the time, obtaining the highest P-state leads to waste of energy as well as thermal problems due to the high CPU temperature. In order to solve these problems, several hardware interfaces are integrated in the CPUs to switch between different frequency/voltage configurations based on the dynamic CPU resource demand. The rapid frequency changes are adjusted through different algorithms depending on the target application~\cite{mallik2006user,choi2002frame,mallik2008picsel}, which is known as CPU performance-frequency scaling.

\noindent\textbf{CPUFreq Subsystem:} The \textit{CPUFreq} subsystem is responsible for the performance scaling of the CPU in a Linux kernel-based operating system. Such subsystem comprises of three defining layers of code named-- the core, the scaling governors, and the scaling drivers. The core setups the common code infrastructure and userspace interfaces as well as defines the layout of the basic framework in which all the components operate. The scaling governors define the scaling algorithm to predict the CPU latency while scaling drivers can access a specific hardware interface to change the P-state based on the request set forth by the scaling governors. Note that, the CPU P-state demonstrates the current frequency and voltage state based on its current workload.

\noindent\textbf{PolicyX Interface:} Initially, the \textit{CPUFreq} core generates a \textit{sysfs} directory named \textit{cpufreq}, under $/sys/devices/system/cpu$ path. Within this directory a policyX sub-directory exists for all of the CPUs associated with the given policy. The policyX directories include policy-specific files to control \textit{CPUFreq} behavior based on the corresponding policy objects. The \textit{CPUFreq} core generates several attributes dependent on the scaling governors and drivers, as well as generic attributes. Some useful attributes under the policyX sub-directory are as follows:
\begin{itemize}
    \item \textit{scaling\_cur\_freq} reports the current frequency of the available CPUs belonging to the specified policy in KHz.
    \item \textit{cpuinfo\_min\_freq} reports the minimum possible operating frequency of the CPU.
    \item \textit{cpuinfo\_max\_freq} reports the maximum possible operating frequency of the CPU.
    \item \textit{cpuinfo\_transition\_latency} reports the time it takes too switch the CPUs from one P-state to another, in nanoseconds.
    \item \textit{scaling\_available\_governors} reports the available scaling algorithms for the defined policy.
    \item \textit{scaling\_governor} reports the scaling algorithm provided by the current driver. This attribute can be changed to other available governors by a root-privileged user.
\end{itemize}

\noindent\textbf{Generic Scaling Governors:} The scaling governors provide different types of frequency scaling algorithms that may be parameterized based on the user demand. The supported scaling governors are as follows: 

\begin{itemize}
    \item \textit{Performance} governor keeps the CPU around the highest frequency, within the scaling\_max\_freq policy limit.
    \item \textit{Powersave} governor keeps the core frequency low when there is no workload still within the scaling\_min\_freq policy limit.
    \item \textit{Userspace} governor allows userspace application to set the CPU frequency for the associated policy by modifying the scaling\_setspeed attribute. The core frequency stays approximately the same all the time.
    \item \textit{Ondemand} governor uses CPU load to determine the CPU frequency selection metric. The estimated CPU load is calculated by the ratio of the non-idle time to the total CPU time. If there are multiple CPUs associated with a policy, the estimated load is computed for all CPUs and the greatest result is taken as the load estimate for the entire policy. The worker routine of \textit{ondemand} governor runs in process context which updates the CPU P-states if necessary. Hence, additional context switches affect its own CPU load metric by increasing the CPU load slightly. 
    The CPU frequencies are set proportional to the estimated load between \textit{cpuinfo\_max\_freq} and \textit{cpuinfo\_min\_freq}.
    \item \textit{Conservative} governor sets the CPU frequency selection metric based on the CPU load. While the CPU load estimation is similar to the \textit{ondemand} governor, the frequency is not changed significantly over short time period. This governor is not suitable for devices with limited power supply capacity such as mobile phones.
    \item \textit{Interactive} governor is designed for latency-sensitive, interactive workloads. Even though the CPU frequency scaling is similar to \textit{ondemand} and \textit{conservative} governors, \textit{interactive} is more aggressive to adjust the frequency compared to other governors. Hence, the issue of under-powering the CPU for a long time is solved with rapid increase in the frequency. This increases the responsiveness of the device as well as improves the battery life on Android devices.
    \item \textit{Schedutil} governor was designed to estimate the load based on the scheduler's Per-Entity Load Tracking (PELT) mechanism. The core frequency is adjusted based on the workload that is scheduled by Completely Fair Scheduler (CFS). This governor is only available on AMD architectures in our experiment setup.
\end{itemize}

\noindent\textbf{Scaling Drivers:} Starting from Sandy Bridge Intel CPUs, Intel P-state driver is introduced for Intel Core CPUs on Linux~\cite{intel_p} for better battery life and performance on modern Linux systems. The scaling algorithm is combined with the hardware capabilities that Intel processors provide; thus, CPU frequency is not limited by the ACPI capabilities~\cite{intel_performance}. The scaling algorithms are fine-tuned for each CPU family to improve both power and performance. AMD architecture family leverages ACPI P-state driver to adjust the core frequency. However, AMD has more options than Intel architectures to enable better power consumption based on the user choice. Since ARM devices generally use Android system, they have specialized frequency scaling driver called \textit{msm}. Similar to AMD, there are several scaling governors that are supported by the ARM devices.

\noindent\textbf{Turbo Boost Technology}: This technology allows both Intel and AMD CPUs to operate on a higher frequency level beyond their maximum threshold on certain conditions, that is commonly referred to as dynamic overclocking.
While it accelerates the CPU performance by approximately 6\% during high workloads such as high-performance game, video editing, and web page rendering, the energy consumption increases around 16\%~\cite{charles2009evaluation}. Such frequency boosting is triggered by hardware for the x86 architecture, and \textit{turbo\_boost} mode is enabled by default for both Intel and AMD architectures. 
However, the exact moment of triggering such boosting is solely decided by the hardware. Moreover, the turbo mode is not always active as the CPU thermal and power budget should not be exceeded. On the other hand, ARM architectures do not provide any Turbo Boost option for the Android users.

\section{Threat Model}\label{sec:threat_model}

\begin{figure}[t]
  \centering
  \includegraphics[width=\linewidth]{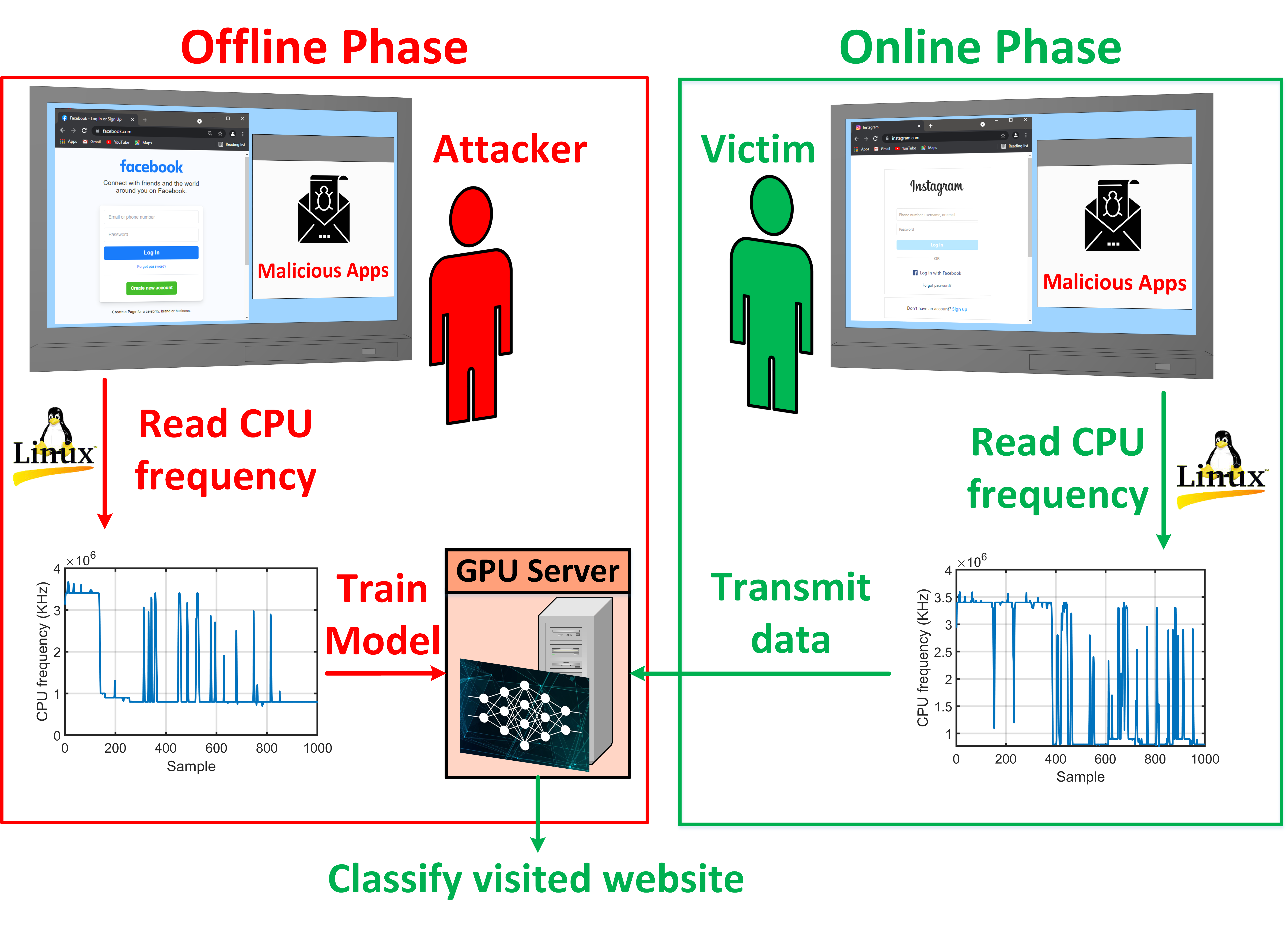}
  \caption{Threat Model for DF-SCA. An attacker can implement a cross-core side-channel attack by exploiting the \textit{cpufreq} interface in the victim device.}
  \label{fig:threat_model}
  \vspace{-5mm}
 \end{figure}

DF-SCA has two phases as illustrated in Figure~\ref{fig:threat_model}. In the offline phase, the attacker monitors the dynamic CPU frequency through the \textit{cpufreq} interface from user space for a certain amount of time, enabled by default in Linux OS, during the website rendering in her own system. After all the websites in a website target list are profiled, a multi-class classification model is trained with the collected frequency measurements. This model is kept in the attacker's server to be queried in the online phase. Note that the attacker's own system is expected to have the same specifications with the victim's device as the frequency traces are expected to alter with different configurations described in Section~\ref{subsec:dfs}.

In the online phase, an attacker places a malicious code in a user-space application to monitor the current frequency, that is installed by the victim in her device. We also assume that the victim visits several websites in a browser environment in which the websites can be considered as private information, and the attacker would like to reveal the website by leveraging the CPU frequency obtained from the attacker application. For such exploitation, an assumption is made where the victim's device is only running a particular browser instead of many applications at a time. Since \textit{cpufreq} interface allows all applications to read the current frequency from all the virtual cores, the attacker can implement a cross-core side-channel attack through the current frequency readings. After the attacker collects a single trace during the website rendering, the trace is sent to the attacker's server in which the pre-trained model is located. Finally, the model is queried in the attacker server to classify the visited website. Both online and offline phases are also used for the password detection on the target Android device in Section~\ref{sec:case2_password_detection}.

\section{Experiment Setup}\label{sec:exp_setup}

The website fingerprinting dataset is collected on four different micro-architectures, namely Intel Comet Lake, Intel Tiger Lake, AMD Ryzen 5, and ARM Cortex-A73 (Table~\ref{tab:CPUFreq_attributes}). The first Intel device is a personal laptop, equipped with Comet Lake micro-architecture and CPU model of Intel(R) Core (TM) i7-10610U CPU @ 1.80GHz. The laptop has four cores with hyper-threading enabled. The installed OS on the device is Ubuntu 20.04 LTS with a Linux kernel version of 5.11.0-46-generic in which \textit{cpufreq} subsystem is supported by default. The processor also supports the Intel Turbo Boost feature to dynamically change the frequency based on the workload. In this study, Google Chrome version 85.0.4183.102 and Tor  browser version 10.5.10 based on Firefox ESR version 78.15.0 are used for browsing. The CPU model of the second Intel device is 11\textsuperscript{th} Gen Intel(R) Core (TM) i7-1165G7 @ 2.80GHz with Tiger Lake micro-architecture. Similar to the Intel Comet Lake, this processor also has four cores and supports Intel Turbo Boost features. The installed OS is Ubuntu 20.04.4 LTS with a Linux version of 5.13.0-44-generic. The versions of the targeted browsers installed in this laptop are Google Chrome 101.0.4951.64 and Tor 10.5.10.

The study is extended for AMD device as well. We have collected the dataset from a laptop assembled with AMD Ryzen 5 5500U CPU with Radeon Graphics. The laptop has six cores and 1.7 GHz base frequency with frequency boosting enabled. The OS and the linux version of this device is similar to the Intel Tiger Lake device. Additionally, the processor also supports AMD Turbo Boost feature. The installed versions of both Google Chrome and Tor Browser remain same as Intel Tiger Lake.

The experiments for mobile phones are performed on Galaxy S8 with four ARM Cortex-A53 and four ARM Cortex-A73 cores. The CPU frequency data is collected on the Cortex-A73 core. The installed OS version is Android 9. The websites are visited in Google Chrome with a version 97.0.4692.98. The password and keystroke detection experiments are performed on Bank of America application with a version of 21.11.04.

Both offline and online phases of the model training and interfering for website and password classification tasks are run on a high-end server equipped with an Nvidia GeForce RTX 3090 GPU card. The CPU model of the server is Intel(R) Xeon(R) CPU E5-1650 v4 @ 3.60GHz, which has six cores with two threads per core.

\begin{table}[ht!]
\centering
\small
\caption{The values of the relevant \textit{cpufreq} attributes for Intel, AMD, and ARM devices are given. The given \textit{scaling\_gov} values are the default values.}
\label{tab:CPUFreq_attributes}
\setlength{\tabcolsep}{3pt}
\begin{tabular}{|l|c|c|c|c|} 
\hline
\multirow{2}{*}{\textbf{Attribute}} & \multicolumn{4}{c|}{\textbf{Micro-architecture}}                                                                                                                                                                                                                                                                 \\ 
\cline{2-5}
                                    & \begin{tabular}[c]{@{}c@{}}\textbf{Intel}\\\textbf{Comet Lake}\end{tabular} & \begin{tabular}[c]{@{}c@{}}\textbf{Intel}\\\textbf{Tiger Lake}\end{tabular} & \begin{tabular}[c]{@{}c@{}}\textbf{AMD}\\\textbf{Ryzen 5}\end{tabular} & \begin{tabular}[c]{@{}c@{}}\textbf{ARM}\\\textbf{Cortex- A73}\end{tabular}  \\ 
\hline
\textit{base\_freq}                 & 1.8 GHz                                                                     & 2.8 GHz                                                                     & 1.7 GHz                                                                & N/A                                                                         \\ 
\hline
\textit{max\_freq}                  & 4.9 GHz                                                                     & 4.7 GHz                                                                     & 4.06 GHz                                                              & 2.36 GHz                                                                    \\ 
\hline
\textit{min\_freq}                  & 0.4 GHz                                                                     & 0.4 GHz                                                                     & 1.4 GHz                                                                & 0.8 GHz                                                                     \\ 
\hline
\textit{scaling\_driv}              & intel\_pstate                                                               & intel\_pstate                                                               & acpi-cpufreq                                                           & msm                                                                         \\ 
\hline
\textit{scaling\_gov}               & powersave                                                                   & powersave                                                                   & ondemand                                                               & interactive                                                                 \\ 
\hline
\textit{turbo\_boost}               & $\checkmark$                                                                & $\checkmark$                                                                & $\checkmark$                                                           & N/A                                                                         \\
\hline
\end{tabular}
\end{table}
\section{Case Study 1: Website Detection}\label{sec:case1}

In the first case study, we perform website fingerprinting attacks on Intel, AMD, and ARM architectures. As described in Section~\ref{subsec:dfs}, there are several attributes under PolicyX sub-directories which are produced by the $CPUFreq$ core, and some of these attributes affect the current CPU frequency readings. In this study, the relevant attributes for the victim's devices are listed in Table \ref{tab:CPUFreq_attributes}. Initially, we proceed with the default scaling governor for the individual devices. The default scaling governors for Intel, AMD, and ARM devices used in this study are \textit{powersave}, \textit{ondemand}, and \textit{interactive}, respectively. The impacts of different scaling governors will be explored in Section~\ref{sec:discussion}. The Turbo Boost feature for both Intel and AMD devices are enabled by default, which eventually improves the system responsiveness for rapid system workload variations during the data collection. The \textit{scaling\_cur\_freq} attribute through \textit{cpufreq} interface is utilized to collect real-time CPU frequency for each core that provides the current CPU frequency to userspace applications in the system.

The website fingerprinting attack is motivated by the quick responses of Intel, AMD, and ARM architectures to the rapidly changing workload in browsers during web page rendering. As our threat model accounts for a cross-core attack, we assume that the target browser is running in a specific core. We assume that the core frequency of the browser's assigned core dynamically changes while the browser is rendering the content, which leads to distinguishable fingerprints for each website.
It is to be noted that, the range of dynamic frequency during different workloads always stay within the range of \textit{max\_freq} and \textit{min\_freq} as specified in Table~\ref{tab:CPUFreq_attributes}. 

\begin{figure}[t!]
  \centering
  \includegraphics[width=\linewidth]{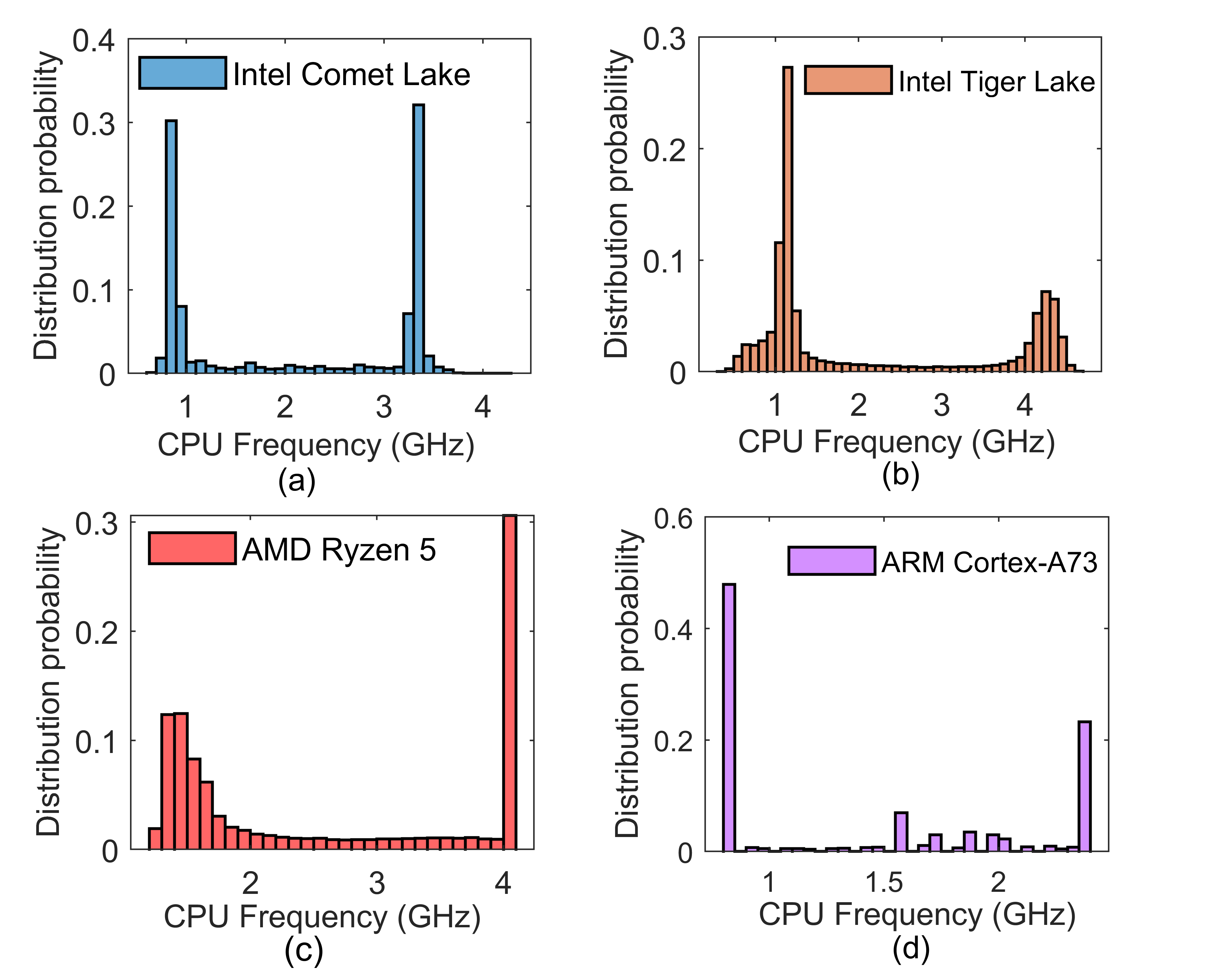}
  \caption{The distribution of CPU frequency readings over 100 websites collected from (a) Intel Comet Lake, (b) Intel Tiger Lake, (c) AMD Ryzen 5, and (d) ARM Cortex-A73.}
  \label{fig:histogram}
  \vspace{-5mm}
\end{figure}

In Figure~\ref{fig:histogram}, the CPU frequency distributions for four different architectures are illustrated, in which the data set is collected from 100 different websites in the Google Chrome browser. The core frequency for the Intel Tiger Lake, AMD Ryzen 5, and ARM Cortex A73 architectures is mostly close to the \textit{max\_freq} or \textit{min\_freq}. On the other hand, for the Intel Comet Lake architecture, the highest CPU frequency can reach up to 3.6 GHz, which is considerably lower than the given maximum frequency. This shows that the dynamic frequency values for Intel Comet Lake are in a comparatively narrow range compared to other architectures, e.g., the actual core frequency alters between 800 MHz and 3.6 GHz as shown in Figure~\ref{fig:histogram}. Although the allowed maximum frequency is 4.9 GHz in Intel Comet Lake, Intel Turbo Boost mechanism increases the frequency up to 3.6 GHz for website rendering due to the thermal and energy consumption limits. In contrast, the AMD Ryzen 5 architecture operates at the highest possible core frequency when the workload is high on the CPU, which is comparatively different than Intel architectures.

\subsection{Data Collection}

While opening a website in a browser, first, essential components of a website are downloaded over network such as HTML file, cascading style sheets (CSS), and Javascript (JS) scripts as well as images. The web page layout is rendered on the screen with the necessary contents for displaying. Hence, during a website rendering, a browser utilizes several computational resources, e.g., CPU, memory, and GPU. DF-SCA only targets the CPU utilization of the browser as the \textit{cpufreq} interface only exposes the CPU frequency to userspace applications.

For the data collection, Algorithm~\ref{alg:data_collection} is used with different parameters that can be tuned for the website fingerprinting \footnote{The dataset and the code will be made available in GitHub: \url{https://github.com/Diptakuet/DF-SCA-Dynamic-Frequency-Side-Channel-Attacks-are-Practical.git}}. Our target browsers are Google Chrome and Tor browsers. In our attack scenario, $N_s = 1000$ samples for each website are collected with an interval of $T_i = 10$ ms for Google-Chrome browser. The number of samples is increased up to 3000 for the Tor browser, as it takes comparatively more time to load a website. In total, $N_M=100$ measurements are collected for each website to evaluate the performance of the website fingerprinting.  100 websites are monitored which are selected from Alexa top 500 sites and listed in Appendix, Table~\ref{tab:google_websites_Intel}.

Figure~\ref{fig:data_analysis} illustrates example website fingerprints based on frequency readings of three websites for the Google Chrome browser recorded from the Intel Comet Lake architecture. Each website has a distinct pattern as the contents of these websites include different JS scripts, images, HTML documents, and plug-in objects. Hence, when they are loaded on the screen, the CPU workload generates a unique fingerprint on the frequency readings. On the other hand, a common pattern exists while visiting the same websites for multiple measurements. Although there are noisy samples due to the background noise, it is expected that advanced machine learning models can classify different websites based on the frequency measurements. 

The same algorithm is utilized to collect data for AMD Ryzen 5 micro-architecture. Unlike Intel Comet Lake and Intel Tiger Lake, the AMD Ryzen 5 has six available governors, namely \textit{ondemand, powersave, performance, userspace, conservative,} and \textit{schedutil}. The default scaling governor is \textit{ondemand} with acpi-cpufreq scaling driver. The range between the \textit{max\_freq} and \textit{min\_freq} limit is comparatively narrower than the Intel devices (Table~\ref{tab:CPUFreq_attributes}). The minimum CPU frequency limit is 1.4 GHz, which is 3.5 times higher than Intel Comet Lake and Tiger Lake micro-architecture.

\begin{algorithm}[t!]
\caption{Data Collection Algorithm for Each Website}\label{alg:data_collection}
\tcp{$T_i$ is the interval between each readings}
\tcp{$N_s$ is the number of samples}
\tcp{$N_m$ is the number of measurements per website}
\tcp{$url$ is the web-page address}
\tcp{$f$ is the CPU frequency}
\KwInput{$T_i,N_s,N_M,url$}
\KwOutput{$f$}
\For{$i \gets 1$ to $N_M$}{
    Run $url$ in the browser \;
    \For{$j \gets 1$ to $N_s$}{
          $f \gets$ Read $scaling\_cur\_freq$ \;
          sleep $T_i$ ;
          }
    Close the browser \;
    sleep $1s$ \;
}
\end{algorithm}

For the ARM-based experiments, we focus on the utilization of A-73 cores as the front-end user activity is generally handled by the big cores. The CPU frequency can be accessed in the same way as in the previous scenario since Android makes use of the Linux kernel. However, the attributes on Android are significantly different than Intel devices as given in Table~\ref{tab:CPUFreq_attributes}. There are six supported governors on Android devices, namely \textit{interactive}, \textit{conservative}, \textit{ondemand}, \textit{userspace}, \textit{powersave}, and \textit{performance}. In our device, the scaling governor is \textit{interactive} by default, which maximizes the responsiveness of the CPU with rapid changes in the CPU frequency.

The minimum and maximum frequency values on Samsung Galaxy S8 device for A-73 cores are 800 MHz and 2361 MHz, respectively. In total, 23 different frequency values are supported in the range of 800 MHz and 2361 MHz by the system. We observed that even though there are several frequency values are supported, the device operates at 806 MHz or 2361 MHz most of the time in the \textit{interactive} mode. The OS evenly distributes the workload among A-73 cores as the frequency readings for all big cores are always the same. Hence, monitoring one big core during the data collection is sufficient to perform DF-SCA on Android devices. The \textit{boostpulse\_duration} attribute is set to 80 ms which indicates that the CPU frequency of all big cores are set to \textit{hispeed\_freq} value for at least 80 ms when \textit{boostpulse} is set to 1. This effect is further observed in the keystroke detection attack in Section~\ref{sec:case2_password_detection}.

\begin{figure}[t!]
  \centering
  \includegraphics[width=\linewidth]{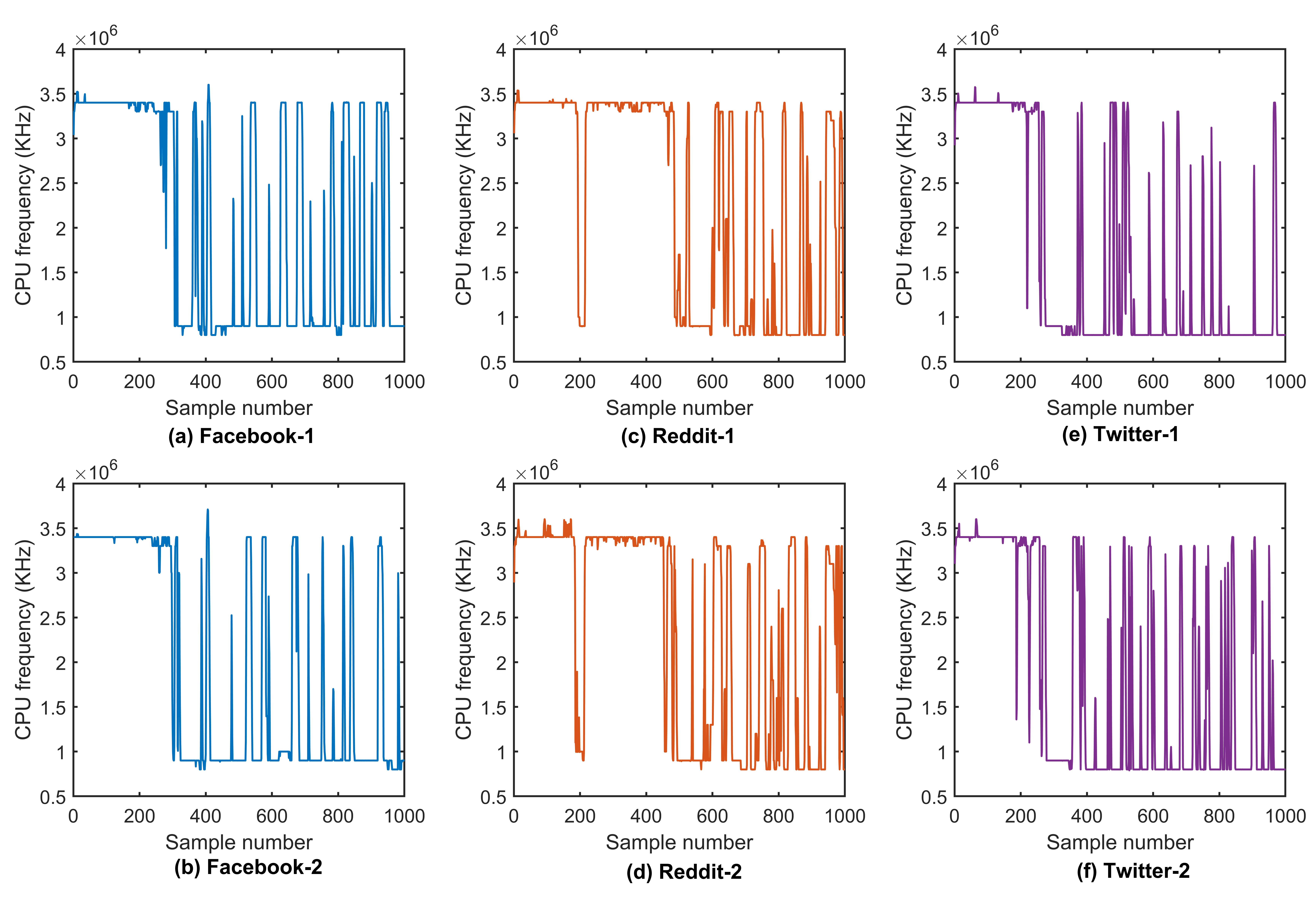}
  \caption{The collected CPU frequency readings while visiting (a) Facebook, (b) Reddit, and (c) Twitter websites on Google Chrome browser. Each website is visited twice, and 1000 frequency measurements are collected with 10 ms interval, i.e., the first 10 seconds of website rendering is monitored.}
  \label{fig:data_analysis}
\end{figure}

\begin{figure}[t!]
  \centering
  \includegraphics[width=\linewidth]{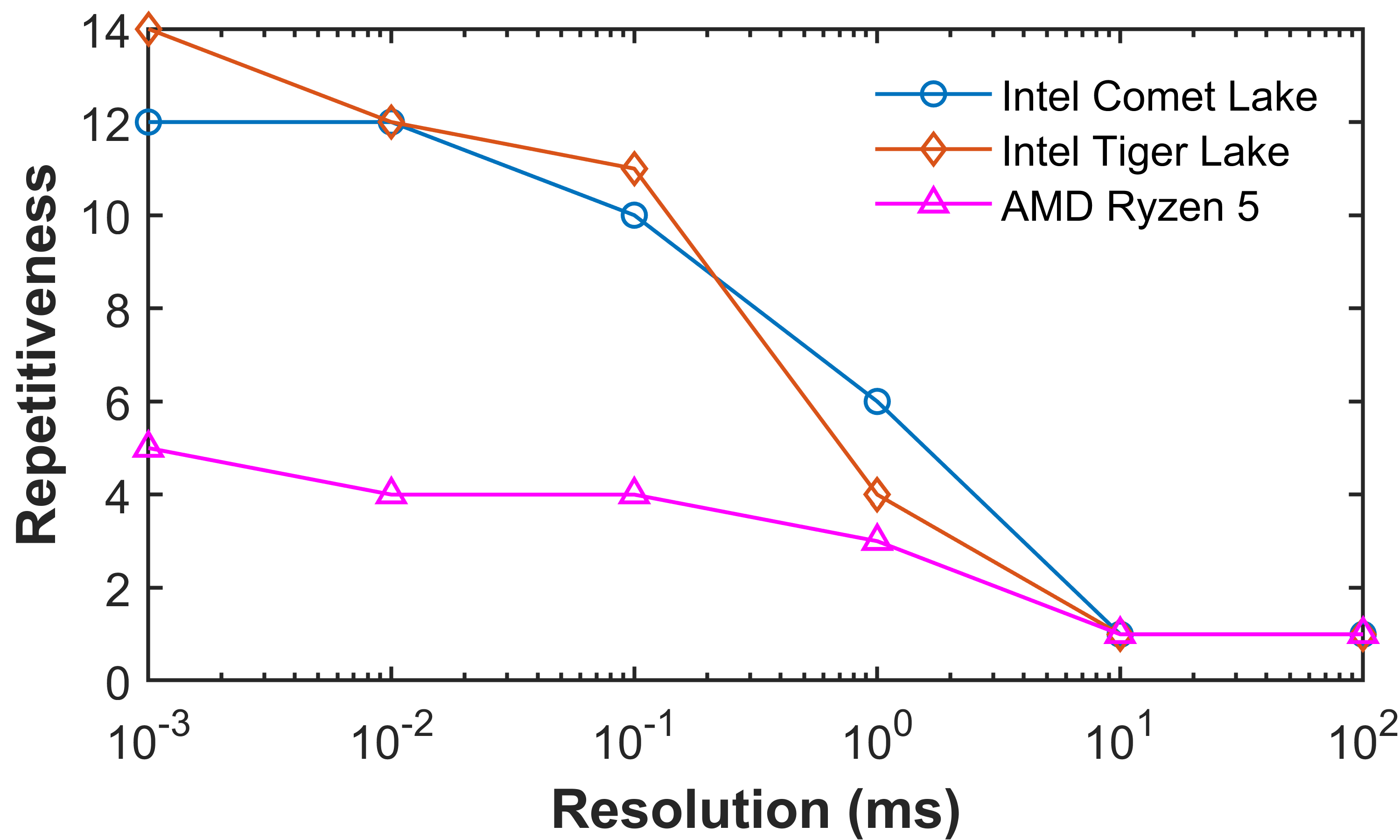}
  \caption{The repeated consecutive CPU frequency readings with different resolutions for the Intel and AMD devices. The highest resolution without the repeated frequency readings is achieved with 10 ms. It is to be noted that repetitiveness as 1 refers to unique readings, i.e., appears once after each consecutive delay.}
  \label{fig:resolution_repetitiveness}
  \vspace{-5mm}
 \end{figure}

\noindent\textbf{\textit{cpufreq} Reading Resolution:} The resolution of DF-SCA has a significant role on the website fingerprinting success rate since higher resolution enables attackers to capture a more detailed fingerprint. In order to evaluate the resolution limit, certain amount of delay is placed between each frequency reading (line 5 in Algorithm~\ref{alg:data_collection}). We observe that the number of repeated values increases with the decreasing amount of delay between each reading, which leads to the optimal delay as 10ms for Intel and AMD architectures as shown in Figure~\ref{fig:resolution_repetitiveness}. The speed of querying the \textit{cpufreq} interface on Android devices is different than Intel and AMD architectures. This value is defined by the \textit{min\_sample\_time} in the \textit{interactive} governor, which is set to 20 ms by default. Hence, the maximum resolution that we can achieve on our Android device is 20 ms. This sampling rate is twice slower than Intel and AMD architectures. On the other hand, the background noise is significantly lower than other architectures as a big portion of the background applications is put in the sleep mode. Since the resolution is already limited to a lower value by the \textit{min\_sample\_time} attribute, no further investigation is performed on the frequency sampling rate. 

\subsection{Google Chrome Website Fingerprinting}\label{subsec:google_chrome_fingerprint} 

We initiated our website fingerprinting attack by targeting Google Chrome browser due to its world-wide usage. The experiment is carried on Intel, AMD, and ARM devices described in Section~\ref{sec:exp_setup}. A data set consists of 100 measurements for 100 different websites is collected based on the data collection mechanism given in Algorithm~\ref{alg:data_collection}. As each measurement consists of 1000 samples, the overall size of the data set is $10000 \times 1000$, where the dimension of a feature vector is $1 \times 1000$ for each website. Since the measurements carry website-specific information in time-series format, one-dimensional CNN (1D-CNN) models are leveraged to learn temporal relations in the measurements. The proposed CNN architecture along with the relevant defining parameters is listed in Appendix, Table~\ref{tab:CNN_arch_1}. The model has four convolutional layers, two maximum pooling layers, two dropout layers, and three dense layers. For the offline scenario, 80\% of the collected data set is used to train the CNN model while 10\% of the data set is given as validation data. The hyper-parameters are adjusted manually to improve both the training and validation accuracy. The \textit{adam} optimizer is leveraged to adapt the learning rate for the objective function while the \textit{categorical cross-entropy} is used to compute the training and validation loss. The CNN model is trained for 50 epochs in the GPU server which takes around 45 seconds. After achieving a stable validation accuracy, the model is saved to be used in the online phase. The remaining 10\% of the data set is fed into the pre-trained model to evaluate the performance of the model. In order to evaluate the performance of the CNN-based model with other ML algorithms, three additional ML algorithm-based models, namely Support vector machine (SVM), K\textsuperscript{th}-nearest neighbor (KNN), and Random forest (RF) are trained. The acquired test accuracy for each models for the individual CPU models with separate micro-architectures are given in Table~\ref{tab:accuracy}.

The test accuracy with the CNN model for the Intel Comet Lake architecture is 94.5\%. Although both SVM and RF models provide more than 90\% accuracy, the CNN model outperforms others. The highest test accuracy of 97.6\% is obtained for the Intel Tiger Lake architecture among all devices, which is classified by the CNN model. It is to be noted that, the default scaling governor for both Intel architectures is \textit{powsersave}.  

The DF-SCA-based website fingerprinting attack is also applicable on AMD devices as presented in Table~\ref{tab:accuracy}. A separate CNN model is trained on the AMD Ryzen 5 architecture. With the default scaling governor \textit{ondemand}, the CNN model achieves 93.1\% accuracy during the online phase tested on this device. The second highest accuracy is obtained from the SVM model with an accuracy of 95.8\%. Finally, the similar experiment is performed on the ARM Cortex A-73 cores. The test accuracy of the pre-trained CNN model for the ARM device is 87.3\%, which is comparatively lower than the Intel and AMD devices. Note that, the resolution of CPU frequency is limited to 20 ms by default for the ARM device. We believe that the lower resolution compared to Intel and AMD devices decreases the success rate of the website fingerprinting attacks.

\begin{table*}[ht!]
\centering
\setlength{\tabcolsep}{13pt}
\caption{Test accuracy for different setups with their default scaling governor mode explored with four ML models}
\label{tab:accuracy}
\begin{tabular}{|c|c|c|c|c|c|c|} 
\hline
\multirow{2}{*}{\textbf{Micro-architecture}} & \multirow{2}{*}{\textbf{Governor}}  & \multirow{2}{*}{\textbf{Browser}} & \multicolumn{4}{c|}{\textbf{Test Accuracy}}               \\ 
\cline{4-7}
                                             &                                     &                                   & \textbf{CNN} & \textbf{SVM} & \textbf{KNN} & \textbf{RF}  \\ 
\hline
\multirow{3}{*}{Intel Comet Lake}            & \multirow{3}{*}{\textit{powersave}} & Chrome                            & 94.5\%~      & 92.0\%~      & 74.6\%~      & 93.7\%~      \\ 
\cline{3-7}
                                             &                                     & Tor                               & 73.7\%~      & 64.9\%~      & 33.6\%~      & 63.6\%~      \\ 
\cline{3-7}
                                             &                                     & Tor (Top 5 score)                 & 93.0\%~      & 86.6\%~      & 54.0\%~      & 86.2\%~      \\ 
\hline
\multirow{3}{*}{Intel Tiger Lake}            & \multirow{3}{*}{\textit{powersave}} & Chrome                            & 97.6\%~      & 95.8\%~      & 84.3\%~      & 93.0\%~      \\ 
\cline{3-7}
                                             &                                     & Tor                               & 68.7\%~      & 51.9\%~      & 16.2\%~      & 30.4\%~      \\ 
\cline{3-7}
                                             &                                     & Tor (Top 5 score)                 & 86.1\%~      & 78.7\%~      & 30.9\%~      & 55.0\%~      \\ 
\hline
\multirow{3}{*}{AMD Ryzen 5}                 & \multirow{3}{*}{\textit{ondemand}}  & Chrome                            & 93.1\%~      & 90.4\%~      & 78.4\%~      & 84.9\%~      \\ 
\cline{3-7}
                                             &                                     & Tor                               & 60.3\%~      & 50.8\%~      & 24.7\%~      & 29.8\%~      \\ 
\cline{3-7}
                                             &                                     & Tor (Top 5 score)                 & 87.0\%~      & 83.2\%~      & 46.5\%~      & 58.2\%~      \\ 
\hline
ARM Cortex-A73                               & \textit{interactive}                & Chrome                            & 87.3\%~      & 71.7\%~      & 38.6\%~      & 69.6\%~      \\
\hline
\end{tabular}
\end{table*}

\subsection{Tor Browser Website Fingerprinting:} \label{subsec:tor_browser}

In this scenario, it is anticipated that the DF-SCA is capable of tracking the victim's activities even while using the Tor browser. Again, the same data collection strategy is applied; however, the run-time is prolonged from 10 seconds to 30 seconds, as the Tor Browser has launch overhead and network delays while loading websites. Hence, the similar program as shown in Algorithm~\ref{alg:data_collection} is also followed for the Tor browser scenario with small modifications, such as the number of samples, $N_s$, is set to 3000 in this case.

In the data collection, 100 different websites for the Tor browser scenario are recorded. Each website is run 100 times to prepare a datas et that is adequate to train a CNN model. The targeted website list is changed since the Tor browser is mostly used by users who are not willing to their identity to websites and Internet Service Providers (ISPs). Hence, several whistle-blowing websites are added to the list while restricted websites in the Tor browser are excluded as given in Appendix, Table~\ref{tab:websites_tor}. The size of the entire data set for the Tor browser is $10000 \times 3000$. For each measurement, the size of the feature vector is $1 \times 3000$ in this case. While the architecture of the CNN model is almost the same as the Google Chrome scenario, hyper-parameter values are fine-tuned to obtain a higher accuracy as listed in Table \ref{tab:CNN_arch_1}.

The data set is split into 80\%, 10\%, and 10\% for the train, validation, and test data sets, respectively. The simulation time to train the model on the GPU server takes around 273.82 seconds for 50 epochs, which is comparatively longer than the previous scenario as the number of features in each measurement is increased three times. The test accuracy obtained from the CNN model over different CPU models and micro-architectures are summarized in Table~\ref{tab:accuracy}. The test accuracy decreases in Tor browser compared to the Google Chrome browser as the order of the network packets and rendering process significantly change between each measurement. Hence, the fingerprints are more noisy and consequently more difficult to classify in Tor browser. The same adverse effect of Tor browser due to the extensive noise on the classification rate is also observed in other studies~\cite{gulmezoglu2017perfweb,zhang2021red,oren2015spy}. Nevertheless, 73.7\% accuracy for the first website guess indicates that malicious userspace applications can distinguish the websites on privacy-oriented browsers through the \textit{cpufreq} interface on the Intel Comet Lake architecture. For Intel Tiger Lake and AMD Ryzen 5 architectures, the highest test accuracy drops to 68.7\% and 60.3\%, respectively. Interestingly, the KNN models cannot learn website-specific fingerprints on the Tor browser scenario. Although the test accuracy on the Tor browser is comparatively lower than the Google Chrome scenario, we also provide the top 5 accuracy for Tor browser scenario, which corresponds to the accuracy rate at which the correct website belongs among the top 5 predictions of the ML model. As presented in Table~\ref{tab:accuracy}, the highest top 5 score for Intel Comet Lake is 93\% that is obtained from the CNN model. For the Intel Tiger Lake and AMD Ryzen 5 architectures, the top 5 scores are 86.1\% and 87\%, respectively. It can be observed that the CNN-based models outperform SVM, KNN, and RF-based models for both Google Chrome and Tor browser scenarios.

\section{Case Study 2: Password Detection}\label{sec:case2_password_detection}

In the password detection scenario, we assume that a phone user enters her password to log into her account in a banking application. Our goal is not to outperform the existing works in the keystroke attack literature, but rather demonstrates DF-SCA attack has sufficient resolution and accuracy to perform a password detection attack. For the target, Bank of America (BoA) mobile application is chosen. 
 
\begin{figure}[t!]
  \centering
  \includegraphics[width=\linewidth]{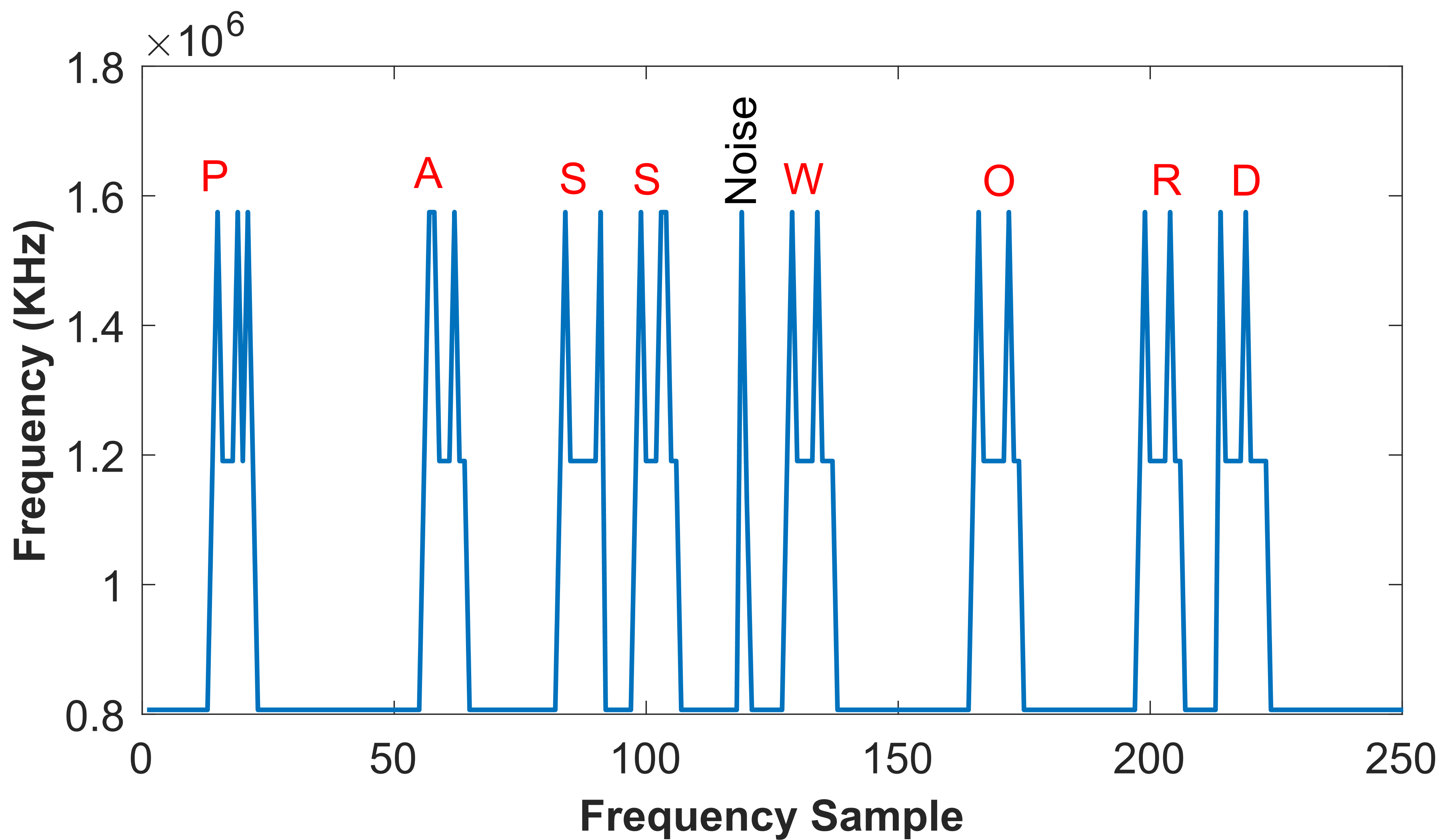}
  \caption{Example keystroke visualization with frequency readings. The data is captured while a user is entering "password" in the BoA application on Galaxy Samsung S8. The CPU frequency does not exceed 1.6 GHz.}
  \label{fig:keystroke_example}
  \vspace{-5mm}
 \end{figure}

First, we focus on distinguishing the user keystrokes from other activities and system noise. For that purpose, a phone user is selected to enter random passwords in the BoA application while a malicious application is reading the current CPU frequency every 20ms. The collected keystrokes in Figure~\ref{fig:keystroke_example} can be identified with three common properties as follows:

\begin{itemize}
    \item A single keystroke length on frequency measurements changes between 8 and 12 samples.
    \item The big cores' frequency increases up to 1.6GHz.
    \item If two consecutive keystrokes are close to each other, the length of a keystroke pattern is higher than 12 samples.
\end{itemize}

Our first observation is that OS can handle the keystrokes without increasing the CPU frequency to the maximum value (2.3 GHz), which eliminates unnecessary power consumption to extend the battery life. It is important to note that after a keystroke is entered, the CPU frequency at most jumps to 1.6GHz. Next, it takes 200 ms in average to decrease the frequency to 800 MHz that is the idle CPU frequency in our mobile phone. Hence, an attacker is able to distinguish the keystrokes that have at least 200 ms between each key press with DF-SCA. When the time difference between two consecutive keystrokes is less than 200ms, the CPU frequency stays above 1.2 GHz with a length of higher than 12 samples. The reason is that the OS detects the user activity, and it keeps the current frequency at least the value specified by the \textit{hispeed\_freq}. If the length of a keystroke is more than 12 samples, then it is assumed that there are two consecutive keystrokes. Even though the attacker does not have enough resolution to measure the time difference between two close keystrokes, detecting the correct number of keystrokes improves the detection rate for the number of keystrokes as well as the password detection accuracy with multiple guesses.

After the necessary properties for keystroke detection are extracted from the frequency readings, 50 out of 200 most used passwords on Internet have been selected~\cite{password_list} to be monitored. The length of the passwords varies from 6 to 9 characters as given in Appendix, Table~\ref{tab:passwords}. Note that, the chosen passwords are not eligible to be valid passwords for the BoA application. However, our purpose is to show that an attacker can create a dataset with the targeted passwords by collecting frequency readings. The phone user entered 50 distinct passwords for at least 10 times while the malicious application monitors the CPU frequency values in the ARM Cortex A-73 cores. In total, 1252 password measurements were collected from 50 distinct passwords. First, the keystroke detection rate is calculated by using three properties of the keystroke patterns. The achieved keystroke detection rate is 95\%. It is observed that the background noise is sometimes treated as a keystroke. Moreover, the third property increased the detection rate from 90\% to 95\% in our case study. This improvement is more visible when the passwords have the same letter next to each other such as 'aa' or '11'.

\begin{figure}[ht!]
  \centering
  \includegraphics[width=\linewidth]{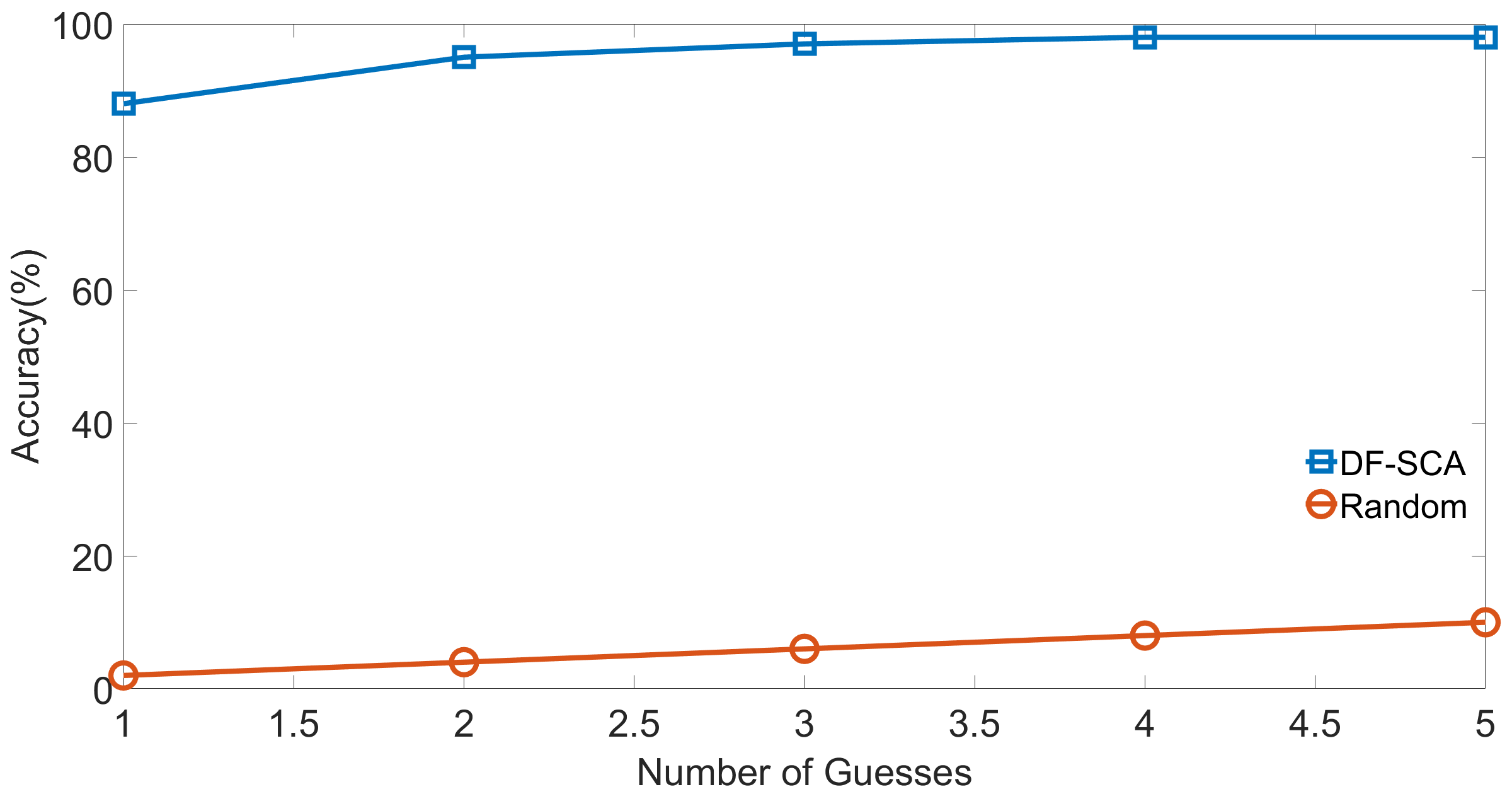}
  \caption{The password detection accuracy with increasing number of guesses. The correct password can be found with 97\% with three guesses among 50 passwords in the BoA application on Galaxy Samsung S8.}
  \label{fig:password_guess}
  \vspace{-5mm}
 \end{figure}

After keystroke timings are detected for each password, the password detection phase starts. The most important information while distinguishing the passwords is the inter-keystroke timings since it provides sufficient information on the location of consecutive keystrokes. It is assumed that if the locations of two subsequent keystrokes are close to each other on the screen, the time difference between the keystroke presses is expected to be low. After the inter-keystroke timings are determined, 10 measurements for each password are selected randomly to prevent the bias in the model. The 70\% of the collected measurements are used to train the KNN model while the rest is used to evaluate the trained model. The model is trained with the euclidean distance and four nearest neighbor parameters. The model can guess the correct password with 88\% success rate with one guess on the test data set, which is 44 times better than a random guess among 50 passwords. The success rate gradually increases with the number of guesses as given in Figure~\ref{fig:password_guess}. With only 3 guesses, the success rate is 97\% which is the number of allowed tries for the BoA application before the account is locked.

\section{Related Work}\label{sec:related_work}

In this section, we give an overview of previous studies on the website fingerprinting and keystroke detection attacks.

\begin{table*}[ht!]
\caption{Previous works based on different side-channel profiling techniques for website fingerprinting. For each work, attack vector, resolution, targeted browser, classification accuracy, and number of websites profiled are given.}
\label{tab:comparison}
\begin{tabular}{|l|l|l|l|l|l|}
\hline
\textbf{Work} & \textbf{Attack Vector} & \textbf{Resolution} & \textbf{Browser} & \textbf{Accuracy (\%)} & \textbf{\# of Websites} \\
\hline
\textbf{DF-SCA} & Frequency scaling   & \textbf{10 ms} & Chrome/Tor  & \textbf{97.6} & 100 \\
\hline
Rendered Insecure~\cite{naghibijouybari2018rendered} &   GPU memory API   & 60 $\mu$s  & Chrome & 90.4 & 200\\
\hline
PerfWeb~\cite{gulmezoglu2017perfweb} &  Performance counters    & 40 $\mu$s    & Chrome/Tor  & 86.4 & 30 \\
\hline
RedAlert~\cite{zhang2021red} & Intel RAPL & 1 ms & Chrome  & 99 & 37\\
\hline
Shusterman et al.~\cite{shusterman2019robust} &   Last-level cache    & 2 ms  & Firefox/Chrome/Tor  & 80 & 100\\
\hline
Spreitzer et al.~\cite{spreitzer2016exploiting} & Data-usage & 20 ms & Tor  & 95 & 100\\
\hline
Zhang et al.~\cite{zhang2018level} & iOS APIs & 1 ms & Safari  & 68.5 & 100\\
\hline
Memento~\cite{jana2012memento} & procfs & 10 $\mu$s & Chrome  & 78 & 100\\
\hline
Loophole~\cite{vila2017loophole} & shared event loop & 25 $\mu$s & Chrome  & 76.7 & 500\\
\hline
\end{tabular}
\end{table*}

\subsection{Website Fingerprinting Attacks}

Website fingerprinting attacks can be implemented by utilizing distinct methods since visited websites take advantage of several technologies such as network connection, browser application, OS, microarchitecture, and so on. If an attacker can monitor certain system-wide or network-based features with sufficient resolution from user-space applications, website fingerprinting attacks become feasible. In order to give an overall insight for website fingerprinting attacks, we cover several techniques:

\noindent\textbf{Network-based Fingerprinting:} The visited websites in a browser require information from servers to transfer the website content to the user device. The transmitted network packets include characteristic information such as length, timing, and protocol, which can be monitored by third-party attackers on the network to detect the visited websites. Since each website has distinct network packet sequence and features, network traffic analysis improves the website fingerprinting attack accuracy even in the encrypted communication~\cite{hintz2002fingerprinting}. Cai et al.~\cite{cai2012touching} demonstrated that defenses against website fingerprinting on Tor browsers are not effective against network traffic analysis. Hayes et al.\cite{hayes2016k} further improved the robustness of website fingerprinting attacks with k-fingerprinting technique. Jansen et al.~\cite{jansen2018inside} applied website fingerprinting from middle relays to enable wide-scale monitoring on Tor network. Li et al.~\cite{li2018measuring} measured the amount of information leaked on the Tor network to evaluate the effectiveness of anonymization on the network. Panchenko et al.~\cite{panchenko2016website} discuss the applicability of website fingerprinting in a real-world scenario by comparing several strategies. Moreover, a large portion of the proposed attacks include a ML/DL algorithm such as Support Vector Machines~\cite{wang2013improved}, Random Forest~\cite{hayes2016k}, Decision Tree~\cite{juarez2014critical}, and Convolutional Neural Network~\cite{sirinam2018deep,bhat2019var} to improve the website fingerprinting success.

\noindent\textbf{Side-channel Website Fingerprinting:} An adversary can also be placed in a user-space application or a malicious website in a browser where a user visits websites regularly. In this threat model, the adversary can collect information through the underlying microarchitecture or OS features that are accessible by user-space applications. In~\cite{vila2017loophole}, shared event loops are exploited to infer the visited websites. Jana et al.~\cite{jana2012memento} also showed that memory footprint reveals the visited websites. Naghibijouybari et al.~\cite{naghibijouybari2018rendered} demonstrated that the amount of allocated GPU memory during a website rendering can also be leveraged to detect the visited website. The microarchitectural side channels such as cache side-channel attacks can also be implemented from Javascript~\cite{shusterman2019robust} environment to infer the visited websites. More interestingly, Cascading Style Sheets and HTML~\cite{shusterman2021prime} can be utilized to detect the visited websites, which bypasses the current Javascript-based defense mechanisms. In addition, user-space accessible OS features can also be exploited such as Android data usage-statistics~\cite{spreitzer2016exploiting}, Linux \textit{perf} interface~\cite{gulmezoglu2017perfweb}, Intel RAPL interface~\cite{zhang2021red} to reveal the visited websites in the user system. In the latest Linux OS releases, both \textit{perf} and RAPL interface access from the userspace applications are disabled by the OS to prevent the user activity tracking.

\vspace{-2mm}

\subsection{Keystroke Recovery}

Other than cryptographic key extraction, and website fingerprinting, attackers leverage software-based side-channel attacks to leak sensitive user input. Several studies showed that the inter-keystroke timings can be monitored with high resolution side-channel attacks to recover bi-grams and tri-grams that lead to high accuracy for password detection attacks. For instance, microarchitectural components reveal the keystrokes through cache activity~\cite{gruss2015cache,ristenpart2009hey} or DRAM usage~\cite{pessl2016drama}. Keystroke recovery can also be implemented through the browser environment by monitoring the event queue~\cite{vila2017loophole,lipp2017practical}. In addition, keystroke timings can be obtained through OS interfaces, such as interrupt statistics \texttt{/proc/interrupts}~\cite{diao2016no,zhang2009peeping}, and stack pointer and instruction pointer \texttt{/proc/stat}. In DF-SCA, we leverage a similar approach by querying the current frequency scaling through \textit{cpufreq} interface to recover the keystrokes. 

\subsection{Comparison of DF-SCA with Previous Attacks}

In this section, we compare our Google-Chrome and Tor browser website fingerprinting attacks with the previous works given in Table~\ref{tab:comparison}. DF-SCA outperforms a big portion of previous works even though the side-channel resolution is lower than other profiling techniques. Moreover, DF-SCA is not mitigated by Intel, AMD, and ARM architectures as well as the underlying OS providers in contrast to previous native scenario attacks such as RedAlert~\cite{zhang2021red} and PerfWeb~\cite{gulmezoglu2017perfweb}. While DF-SCA is expected to be more susceptible to the system noise since any additional workload increases the CPU frequency that alters the side-channel measurements, the experiments are performed in a noisy environment in which background applications necessary for the system are still running. Since the attack resolution is limited by the update interval of the \textit{cpufreq} interface on Android devices the lower success rate is achieved on ARM devices compared to Intel and AMD architectures.

There are two closely related works to DF-SCA, which leverages CPU frequency side-channels. Qin et al.~\cite{qin2018website} proposes power consumption modeling for websites based on the CPU frequency on Android devices. Since the power modeling of dynamic frequency is not an effective attack, their success rate is only 55\% for 20 websites. Their work also has no investigation on Intel and AMD devices. They further claim that their work is not applicable from userspace starting from Android 8. Another recent work, Hertzbleed attack~\cite{wan2022hertzbleed}, discovered that dynamic CPU frequency changes based on the hamming weight and hamming distance of the values in the registers. They verify that Hertzbleed is applicable on both Intel and AMD architectures. However, they claim that their attack can be mitigated if Turbo Boost is disabled. DF-SCA shows that website fingerprinting and keystroke detection attacks can be applicable in the absence of Turbo Boost feature on ARM devices. 

\section{Countermeasures}\label{sec:countermeasures}
In this section, we propose several countermeasures to prevent DF-SCA on Linux-based devices.

\noindent\textbf{Restricting Access Privilege for \textit{cpufreq}:} The easiest way to thwart DF-SCA is to restrict the monitoring of \textit{cpufreq} interface from userspace applications in Linux OS. If the \textit{scaling\_cur\_freq} attribute is solely masked by the OS, there would be other resources that could leak information about the CPU frequency such as \textit{stats} under \textit{cpufreq} interface on Android devices.

\noindent\textbf{Resolution Reduction:} An alternative solution would be to decrease the update interval of the \textit{cpufreq} interface since it would be more difficult to track the CPU frequency in real-time with a lower resolution in which the amount of information leaked by DF-SCA can be diminished significantly. Even though this defense mechanism can still be abused by profiling the websites for a longer time, the attack accuracy can be decreased significantly. As an example, Javascript-based timer resolution on Tor browsers is reduced to 100 ms which decreases the website detection rate to 45\% for 100 websites~\cite{shusterman2019robust}.

\noindent\textbf{Artificial Noise:} To mitigate specific attacks such as website fingerprinting, artificial noise can be introduced by the system to mask the rapid frequency changes in the system. Such an approach has been proposed in~\cite{kohlbrenner2016trusted} to protect the browsers against timing attacks by randomly inserting workloads in the system. Since side-channel analysis takes advantage of Deep Learning algorithms frequently, adversarial obfuscation techniques can also be implemented to fool the Deep Learning models~\cite{gulmezoglu2021xai}. Similarly, keystroke attacks can be eliminated by introducing additional keystrokes to make the distribution more uniform~\cite{schwarz2018keydrown}.

\section{Discussion}\label{sec:discussion}

\noindent\textbf{Scaling Governor Analysis:} Even though we show the applicability of DF-SCA website fingerprinting and keystroke detection attacks on Intel, AMD, and ARM devices with default scaling governors, it is still unclear whether other scaling governors provide sufficient information on the visited websites. Hence, we investigate the effect of different scaling governors on the website fingerprinting accuracy on Intel and AMD devices. Although there are only two supported governors on modern Intel architectures, both AMD and Android-based mobile phones offer more scaling governor options.

In order to evaluate the impact of different scaling governors on the website fingerprinting accuracy, we conducted experiments on the Intel Tiger Lake and AMD Ryzen 5 architectures by changing the scaling governor mode. The website fingerprinting accuracy for the Intel Tiger Lake improves slightly when the scaling governor is changed to \textit{performance} from \textit{powersave} as presented in Table~\ref{tab:accuracy_scaling_governor}. It means that the responsiveness of the device is quicker with the \textit{performance} governor, which eventually captures the changes in the browser workload slightly more efficiently. On the other hand, the AMD Ryzen 5 architecture has more available scaling governor options. We observed that the default scaling governor \textit{ondemand} gives the highest website classification accuracy compared to the other five governors as given in Table~\ref{tab:accuracy_scaling_governor}. The website fingerprinting is still viable with different governors with the lowest accuracy of 68.1\% when the performance governor mode is active. Interestingly, the performance and power governors drop the classification accuracy, which shows that Intel and AMD implement different types of frequency scaling even though the governor names are the same. Although the accuracy drops with \textit{performance}, \textit{powersave}, and \textit{userspace} governors, the accuracy still stays high for \textit{schedutil} and \textit{conservative} governors. Interestingly, the \textit{userspace} governor keeps the core frequency around 1.8GHz all the time, which decreases the variations on the core frequency compared to the other governors. However, depending on the changing workload on the system, small variations are observed on the frequency readings as given in Appendix Figure~\ref{fig:userspace_data}. The high accuracy with \textit{userspace} governor shows that even small variations are sufficient to extract meaningful website fingerprints from the frequency readings.

\begin{table} [ht!]
\centering
\caption{The impacts of different scaling governors on website fingerprinting accuracy for Intel Tiger Lake and AMD Ryzen 5 architectures}
\begin{tabular}{|c|c|c|} 
\hline
\multirow{2}{*}{\textbf{~ Scaling governor}} & \multicolumn{2}{c|}{\textbf{Test Accuracy (\%)}}  \\ 
\cline{2-3}
                                             & \textbf{Intel Tiger Lake} & \textbf{AMD Ryzen 5}  \\ 
\hline
performance                                  & \textbf{97.8}            & 68.1                  \\ 
\hline
powersave                                    & 97.6                     & 75.3                  \\ 
\hline
userspace                                    & N/A                       & 80.1                  \\ 
\hline
ondemand                                     & N/A                       & \textbf{97.6}                  \\ 
\hline
conservative                                 & N/A                       & 96.7                  \\ 
\hline
schedutil                                    & N/A                       & \textbf{97.6}                  \\
\hline
\end{tabular}
\label{tab:accuracy_scaling_governor}
\end{table}

\noindent\textbf{Universal ML Model for different CPU models:}
In the previous experiment, we trained separate ML models for Intel, AMD, and ARM architectures to obtain the highest website fingerprinting accuracy. However, it is still unclear whether it is possible to replace the individual ML models with a universal ML model trained with the CPU frequency data from several micro-architectures. Thus, an attacker can use a combined ML model without requiring to know the exact targeted microarchitecture for website fingerprinting. For this purpose, initially, we combined the CPU frequency data collected with \textit{powersave} governor from both Intel Tiger Lake and Intel Comet Lake architectures to train a universal CNN model and evaluated the performance of the universal model with the test data. For the universal CNN model, we obtained an accuracy of 95.9\% for the Google Chrome browser scenario as given in Table~\ref{tab:combined_ML}. Later, we added the CPU frequency data set from the AMD Ryzen 5 architecture collected from \textit{ondemand} governor and created a universal cross-architecture ML model. The overall accuracy for the test data set eventually drops to 92.3\%. However, this result shows that a universal CNN model can still be utilized to eliminate the assumption of knowing the targeted architecture for high accuracy website fingerprinting even for cross-architecture measurements.

\begin{table}[ht!]
\small
\centering
\setlength{\tabcolsep}{2pt}
\caption{The universal ML Model training and evaluation for Intel Tiger Lake, Intel Comet Lake, and AMD Ryzen 5 architectures}
\begin{tabular}{|c|c|} 
\hline
\textbf{Micro-architecture} & \textbf{Test Accuracy (\%)} \\ 
\hline
Intel Comet Lake
  +~ Intel Tiger Lake                       & 95.9  \\ 
\hline
Intel Comet Lake
  +~  Intel Tiger Lake  + AMD Ryzen 5      & 92.3   \\ 
\hline
\end{tabular}
\label{tab:combined_ML}
\end{table}
\section{Conclusion}\label{sec:conclusion}

DF-SCA demonstrated that dynamic frequency scaling feature in modern systems can be exploited in Linux OS through \textit{cpufreq} interface to reveal the visited websites in Google Chrome and Tor browsers as well as entered user passwords on Android mobile phones. The attacker only needs to collect 10 seconds of the frequency values to detect the websites in Google Chrome browser applicable to Intel, AMD, and ARM devices. Even though DF-SCA's resolution is significantly lower than many previous attacks, it is still possible to detect the visited websites with a high accuracy. Moreover, victim keystrokes can be detected with 95\% success rate which yields to a successful password recovery attack with a simple ML classification. As a result, DF-SCA is a potential threat for all the components that take advantage of DFS technology. DF-SCA shows that efficient countermeasures need to be integrated into Linux kernel to prevent the privacy violation of the users against malicious applications.  The access privilege restriction or artificial noise injection might become fruitful countermeasures against such a threat.

\bibliographystyle{plain}
\bibliography{Ref}
\newpage
\appendix

\section{Appendix}

\begin{figure}[ht!]
  \centering
  \includegraphics[width=\linewidth]{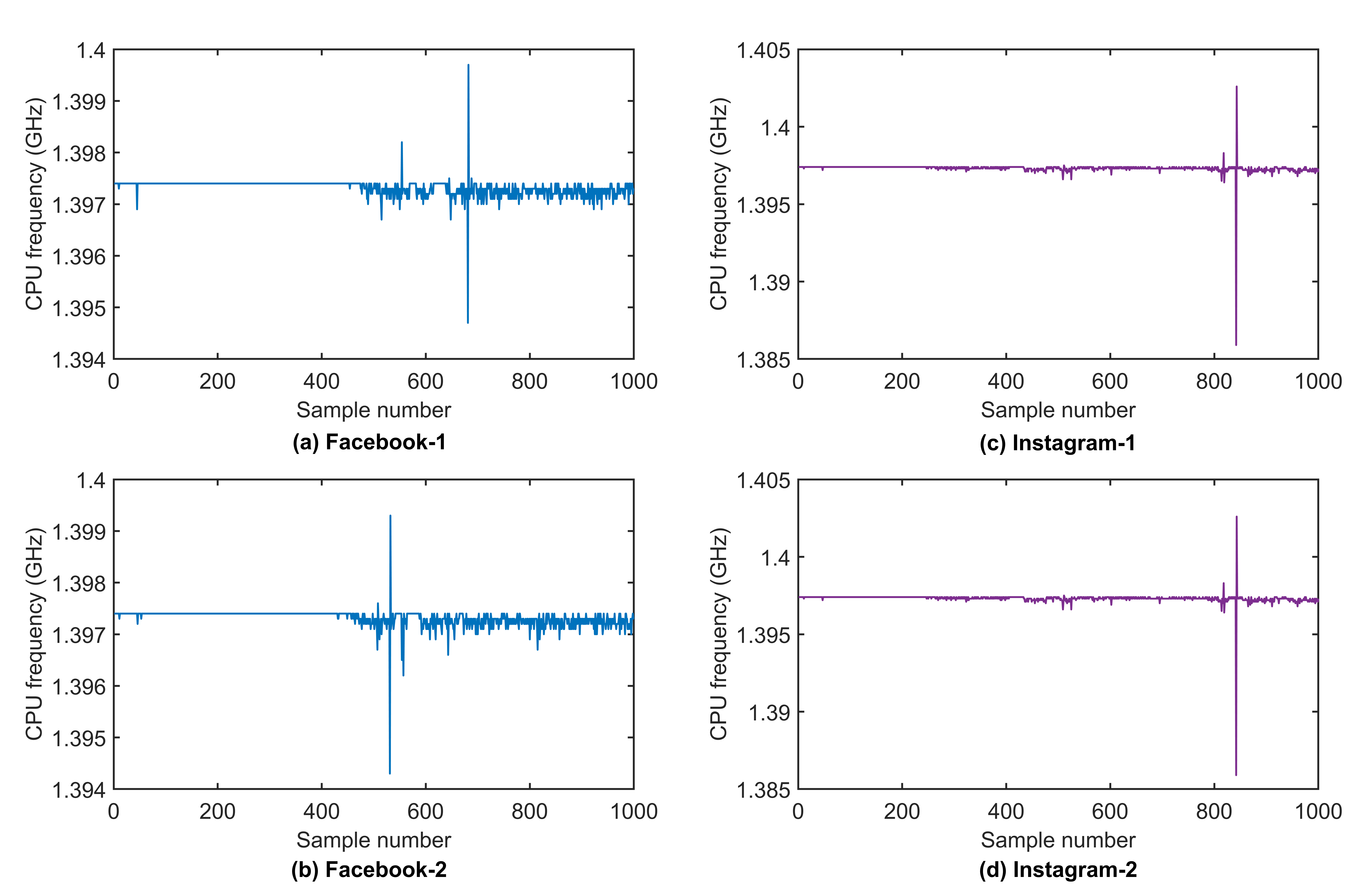}
  \caption{The CPU frequency reading collected from AMD Ryzen 5 with \textit{userspace} governor. Unlike other scaling governors, for \textit{userspace} the CPU frequency does not rise to the maximum frequency limit with different workloads. The frequency variation while browsing Facebook and Instagram web pages is quite low, and interestingly, the frequency remains below its base frequency of 1.4 GHz. Although the variation is quite low, a similar pattern for the same web page can still be noticeable from this figure.}
  \label{fig:userspace_data}
 \end{figure}
 
\newpage
\begin{table}[ht!]
\small
\caption{Profiled Websites for Google-Chrome Scenario}
\label{tab:google_websites_Intel}
\setlength{\tabcolsep}{3pt}
\begin{tabular}{|lll|}
\hline
1. 360.cn                & 36. Taobao.com           & 71. imgur.com        \\
2. Adobe.com             & 37. Target.com           & 72. soundcloud.com    \\
3. Aliexpress.com        & 38. Tiktok.com           & 73. discord.com         \\
4. Amazon.com            & 39. Tmall.com            & 74. booking.com           \\
5. Apple.com             & 40. Twitter.com          & 75. rakuten.com            \\
6. Baidu.com             & 41. Usps.com             & 76. yandex.ru               \\
7. Bestbuy.com           & 42. Weibo.com            & 77. vimeo.com                 \\
8. Blogspot.com          & 43. Wellsfargo.com       & 78. etsy.com                   \\
9. Canva.com             & 44. Hulu.com             & 79. slideshare.net           \\
10. Chase.com            & 45. Ups.com              & 80. cloudfare.com            \\
11. Csdn.net             & 46. Xinhuanet.com        & 81. vice.com               \\
12. Diply.com            & 47. Yahoo.com            & 82. foxnews.com         \\
13. Ebay.com             & 48. Zhanqi.tv            & 83. 9gag.com          \\
14. Facebook.com         & 49. Zillow.com           & 84. slack.com           \\
15. Force.com            & 50. Zoom.com             & 85. airbnb.com            \\
16. Google.com           & 51. Instagram.com        & 86. telegram.org            \\
17. Huanqiu.com          & 52. whatsapp.com         & 87. forbes.com               \\
18. Indeed.com           & 53. bing.com             & 88. roblox.com                \\
19. Intuit.com           & 54. fandom.com           & 89. globo.com              \\
20. Jd.com               & 56. twitch.tv            & 90. w3schools.com         \\
21. Linkedin.com         & 56. twitch.tv            & 91. flipkart.com         \\
22. Live.com             & 57. walmart.com          & 92. bankofamerica.com  \\
23. Msn.com              & 58. imdb.com             & 93. fedex.com         \\
24. Naver.com            & 59. espn.com             & 94. scribd.com         \\
25. Netflix.com          & 60. github.com           & 95. mediafire.com        \\
26. Ntdtv.com            & 61. stackoverflow.com    & 96. researchgate.net     \\
27. Nytimes.com          & 62. douban.com           & 97. softonic.com          \\
28. Office.com           & 63. youtube.com          & 98. rambler.ru            \\
29. Okezone.com          & 64. mail.ru              & 99. washingtonpost.com    \\
30. Paypal.com           & 65. quora.com            & 100. theguardian.com      \\
31. Pinterest.com        & 66. tumblr.com           &                          \\
32. Qq.com               & 67. bbc.com              &                        \\
33. Reddit.com           & 68. medium.com           &                       \\
34. Sohu.com             & 69. dropbox.com          &                      \\
35. Spotify.com          & 70. godaddy.com          &                    \\
\hline
\end{tabular}
\end{table}

\newpage
\begin{table}[ht!]
\small
\caption{Profiled Websites for Tor browser Scenario}
\label{tab:websites_tor}
\setlength{\tabcolsep}{2pt}
\begin{tabular}{|lll|}
\hline
1. Facebook.com         & 36. Publeaks.nl           & 71. yandex.ru         \\
2. Amazon.com           & 37. Aljazeera.com         & 72. vimeo.com         \\
3. Google.com           & 38. Apple.com             & 73. etsy.com          \\
4. Netflix.com          & 39. Adobe.com             & 74. slideshare.net    \\
5. Yahoo.us             & 40. Canva.com             & 75. vice.com          \\
6. Wikipedia.org        & 41. Indeed.com            & 76. foxnews.com       \\
7. Aliexpress.com       & 42. Intuit.com            & 77. 9gag.com          \\
8. Bing.com             & 43. Jd.com                & 78. slack.com         \\
9. Ebay.com             & 44. Qq.com                & 79. telegram.org      \\
10. Reddit.com          & 45. Target.com            & 80. forbes.com        \\
11. Twitter.com         & 46. Tiktok.com            & 81. roblox.com        \\
12. Linkedin.com        & 47. Wellsfargo.com        & 82. globo.com         \\
13. Live.com            & 48. Ups.com               & 83. w3schools.com     \\
14. Diply.com           & 49. Zillow.com            & 84. flipkart.com      \\
15. Ntd.tv              & 50. Zoom.com              & 85. fedex.com         \\
16. Cnn.com             & 51. instagram.com         & 86. scribd.com        \\
17. Pinterest.com       & 52. whatsapp.com          & 87. mediafire.com     \\
18. Office.com          & 53. fandom.com            & 88. researchgate.net  \\
19. Microsoft.com       & 54. instructure.com       & 89. softonic.com      \\
20. Chase.com           & 55. twitch.tv             & 90. rambler.ru        \\
21. Nytimes.com         & 56. walmart.com           & 91. washingtonpost.com \\
22. Blogspot.com        & 57. github.com            & 92. theguardian.com   \\
23. Paypal.com          & 58. stackoverflow.com     & 93. cloudflare.com    \\
24. Wordpress.com       & 59. douban.com            & 94. wordpress.org     \\
25. Espn.com            & 60. mail.ru               & 95. gravatar.com      \\
26. Wikia.com           & 61. quora.com             & 96. brandbucket.com   \\
27. Wikileaks.org       & 62. tumblr.com            & 97. who.int           \\
28. Imdb.com            & 63. bbc.com               & 98. dailymotion.com   \\
29. Balkanleaks.eu      & 64. medium.com            & 99. nature.com        \\
30. Unileaks.org        & 65. dropbox.com           & 100. time.com         \\
31. Globaleaks.com      & 66. godaddy.com           &                       \\
32. Liveleak.com        & 67. imgur.com             &                       \\
33. Globalwitness.org   & 68. soundcloud.com        &                       \\
34. Wikispooks.com      & 69. discord.com           &                       \\
35. Officeleaks.com     & 70. booking.com           &                       \\
\hline
\end{tabular}
\end{table}

\newpage

\begin{table}[ht!]
\centering
\caption{The specifies parameters of the different ML-based models exploited for website classification while considering both google-chrome and tor browser scenario.}
\label{tab:CNN_arch_1}
\setlength{\tabcolsep}{8pt}
\begin{tabular}{|l|l|l|} 
\hline
\multicolumn{2}{|l|}{\textbf{ML-Model~}}       & Attributes                                                                           \\ 
\hline
\multirow{11}{*}{\textbf{CNN}} & Conv1         & \begin{tabular}[c]{@{}l@{}}No. of filters = 64\\ Kernel size = 3\\\end{tabular}      \\ 
\cline{2-3}
                               & Conv2         & \begin{tabular}[c]{@{}l@{}}No. of filters = 64\\ Kernel size = 3\end{tabular}        \\ 
\cline{2-3}
                               & Max Pooling 1 & Pool size = 3                                                                        \\ 
\cline{2-3}
                               & Conv3         & \begin{tabular}[c]{@{}l@{}}No. of filters = 128\\ Kernel size =3\end{tabular}        \\ 
\cline{2-3}
                               & Conv4         & \begin{tabular}[c]{@{}l@{}}No. of filters = 128\\ Kernel size = 3\end{tabular}       \\ 
\cline{2-3}
                               & Max Pooling 2 & Pool size = 3                                                                        \\ 
\cline{2-3}
                               & Dropout       & Rate = 0.5                                                                           \\ 
\cline{2-3}
                               & Dense1        & Units = 256                                                                          \\ 
\cline{2-3}
                               & Dropout       & Rate = 0.5                                                                           \\ 
\cline{2-3}
                               & Dense2        & Units = 128                                                                          \\ 
\cline{2-3}
                               & Dense3        & Units = 100                                                                          \\ 
\hline
\textbf{SVM}                   & \multicolumn{2}{l|}{\begin{tabular}[c]{@{}l@{}}kernel = ’rbf’\\Kernel coefficient, $gamma$ = 'auto'\\Regularization parameter, $C = 3$\end{tabular}}     \\ 
\hline
\textbf{KNN}                   & \multicolumn{2}{l|}{\begin{tabular}[c]{@{}l@{}}No. of neighbors = 5\\Distance metric = ’minkowski’\end{tabular}}  \\ 
\hline
\textbf{RF}                    & \multicolumn{2}{l|}{\begin{tabular}[c]{@{}l@{}}max\_depth = 18\\random\_state = 0\end{tabular}}      \\
\hline
\end{tabular}
\end{table}

\begin{table}[ht!]
\centering
\small
\caption{Profiled Passwords for Password Detection on Android chosen from~\cite{password_list}.}
\label{tab:passwords}
\setlength{\tabcolsep}{4pt}
\begin{tabular}{|llll|}
\hline
\textbf{6 characters}   & \textbf{7 characters}   & \textbf{8 characters}    & \textbf{9 characters} \\
\hline
1q2w3e   & 1234567   & 12345678    & 123456789  \\
123456     & charlie  & password  & password1  \\
123123     & jessica      & myspace1     & target123    \\
qwerty   & samsung    & football  & asdasd123     \\
abc123      & michael        & princess       & liverpool  \\
dragon  & pokemon       & sunshine      & iloveyou1  \\
monkey & welcome & computer  & princess1  \\
daniel       & anthony    &abcd1234       &football1       \\
shadow      & letmein & superman & chocolate   \\
killer    & freedom     & baseball   & qazwsxedc     \\
soccer    &    & michelle   &    \\
tinkle    &     & jennifer   &     \\
jordan    &     &   &     \\
thomas    &    &   &    \\
andrew    &      &  &   \\
hunter    &      &    &    \\
naruto    &     &   &    \\
justin    &     &   &     \\
\hline
\end{tabular}
\end{table}


\end{document}